\documentclass[a4paper,final]{article}

\usepackage{lmodern}
\usepackage[T1]{fontenc}
\usepackage[utf8]{inputenc}
\usepackage{amsmath,physics,slashed}
\usepackage{amsfonts}
\usepackage{mathtools}
\usepackage{amssymb,bbm}
\usepackage{dsfont}
\usepackage{subcaption}
\usepackage{braket}
\usepackage{amsthm}
\usepackage{authblk}
\usepackage{xcolor}
\usepackage{cite}

\usepackage[hmargin=.15\paperwidth,vmargin=.15\paperheight]{geometry}
\usepackage[linktoc=all,colorlinks,allcolors=blue!75!black]{hyperref}

\newcommand{\nc}{\newcommand}
\nc{\rnc}{\renewcommand} 

\rnc{\a}{\alpha}
\rnc{\b}{\beta}
\nc{\g}{\gamma}
\rnc{\d}{\delta}
\nc{\e}{\epsilon}
\nc{\ee}{\varepsilon}
\nc{\z}{\zeta}
\nc{\f}{\phi}
\nc{\m}{\mu}
\nc{\n}{\nu}
\rnc{\r}{\rho}
\rnc{\k}{\kappa}
\rnc{\l}{\lambda}
\nc{\p}{\pi}
\nc{\s}{\sigma}
\rnc{\t}{\tau}
\nc{\w}{\omega}
\nc{\x}{\chi}
\nc{\F}{\Phi}
\rnc{\L}{\Lambda}
\nc{\cL}{\mathcal L}
\nc{\fb}{{\overline f}}
\nc{\zb}{{\bar z}}

\nc{\pd}{\partial}
\newcommand{\eq}[2]{\begin{align}\label{#1}#2\end{align}}

\newcommand{\Tb}{\overline{T}}
\newcommand{\TTb}{T\overline{T}}
\newcommand{\alphab}{\overline{\alpha}}
\newcommand{\Nb}{\overline{N}}

\newcommand{\rf}[1]{(\ref{#1})}
\newcommand{\rfs}[1]{\ref{#1}}
\newcommand{\rt}{\rightarrow} 
\newcommand{\bb}{\overline{b}}
\newcommand{\Fb}{\overline{F}}
\hypersetup{
    pdftitle = {Refining the cutoff 3d gravity / TTbar Correspondence},
    pdfauthor = {Per Kraus, Ruben Monten and Konstantinos Roumpedakis}
}

\title{\Huge Refining the Cutoff 3d Gravity / $\TTb$ Correspondence } 
\date{}
\author[]{Per Kraus}
\author[]{Ruben Monten}
\author[]{Konstantinos Roumpedakis.}
\affil[]{Mani L.\ Bhaumik Institute for Theoretical Physics,  Department of Physics and Astronomy, University of California, Los Angeles, CA 90095, USA}

\begin{document}
\maketitle
\begin{abstract}
    Pure gravity in AdS$_3$ is a theory of boundary excitations, most simply expressed as a constrained free scalar with an improved stress tensor that is needed to match the Brown--Henneaux central charge. Excising a finite part of AdS gives rise to a static gauge Nambu--Goto action for the boundary graviton. We show that this is the $T\Tb$ deformation of the infinite volume theory, as the effect of the improvement term on the deformed action can be absorbed into a field redefinition. The classical gravitational stress tensor is reproduced order by order by the $T\Tb$ trace equation. We calculate the finite volume energy spectrum in static gauge and find that the trace equation imposes sufficient constraints on the ordering ambiguities to guarantee agreement with the light-cone gauge prediction. The correlation functions, however, are not completely fixed by the trace equation. We show how both the gravitational action and the $T\Tb$ deformation allow for finite improvement terms, and we match these to the undetermined total derivative terms in Zamolodchikov's point splitting definition of the $T\Tb$ operator.
\end{abstract}
\pagebreak
{
	\hypersetup{linkcolor=black}
    \tableofcontents
}

\section{Introduction}

The fact that General Relativity involves the dynamics of spacetime geometry implies that the problem of defining  boundary conditions in gravity is bound to be more subtle than in non-gravitational theories.  Gravitational boundary conditions are typically imposed asymptotically, for example at spacelike or null infinity in asymptotically flat spacetimes, or on the conformal boundary of anti-de Sitter spacetime.  It is of interest to ask whether imposing boundary conditions at a ``finite cutoff boundary'' makes sense, for example by imposing a Dirichlet condition on the metric on a non-asymptotic timelike boundary.   One setting where this does seem to make sense --- though not without subtleties --- is the case of pure AdS$_3$ gravity. This example is also of special interest due to its holographic relation to $\TTb$ deformed CFT$_2$ \cite{Smirnov:2016lqw,Cavaglia:2016oda,McGough:2016lol}. 

This problem was studied at both the classical and quantum level in \cite{Kraus:2021cwf,Ebert:2022cle}.  The focus was on the dynamics of the boundary gravitons, since in 3d gravity these are the only propagating degrees of freedom \cite{Brown:1986nw,Banados:1998gg,Maloney:2007ud}. 
  A main result was that  these boundary gravitons are governed by a static gauge Nambu-Goto action defined on the cutoff boundary.   Obtaining this result involved making a highly nontrivial and nonlocal field redefinition between the variables describing the embedding of the boundary surface and those appearing in the Nambu-Goto action.  Quantization of this theory was initiated, revealing the need for renormalization counterterms, which is not surprising given the well-known issues involved in quantizing the square root Nambu-Goto  action. 

In this paper we clarify, consolidate, and extend these results in a way that provides a satisfying agreement between the gravity and  $\TTb$ deformed CFT sides of the story.    We now summarize the main results of our analysis.   Other references on cutoff AdS$_3$ gravity include \cite{Kraus:2018xrn,Cottrell:2018skz,Donnelly:2018bef,Caputa:2019pam,Guica:2019nzm,Donnelly:2019pie,Lewkowycz:2019xse,Mazenc:2019cfg,Llabres:2019jtx,Li:2020pwa,Ouyang:2020rpq,Caputa:2020lpa,Li:2020zjb}.

The general problem involves the classical and quantum theory of boundary gravitons living on top of some background geometry, which could be global AdS, a conical defect, or a BTZ black hole.  We begin by considering the asymptotic AdS boundary case.   The boundary action, which can be obtained either in the Chern-Simons \cite{Coussaert:1995zp,Cotler:2018zff} or metric \cite{Kraus:2021cwf,Ebert:2022cle}  formulation  of 3d gravity, is given by the Alekseev-Shatashvili theory \cite{Alekseev:1988ce,Alekseev:1990mp}.  After a field redefinition, this becomes the standard action for a free scalar, but with a few special features. First, the scalar is not periodic around the boundary cylinder, but obeys a fixed winding condition that encodes the mass and angular momentum of the background gravity solution.  Second, the theory is subject to a gauge symmetry whose origin lies in the isometry group of the background geometry.  Finally, the stress tensor is not canonical but has an improvement term, which is needed in order to realize the Brown-Henneaux central charge.   We carefully go though the quantization of this theory, which has some subtleties depending on the form of the winding condition, showing how to arrive at unitary representations of the Virasoro algebra.  

It is well-known that the Nambu-Goto action is the  $\TTb$ deformed action for an ordinary free scalar with canonical stress tensor \cite{Cavaglia:2016oda}.  As noted above, the scalar arising from gravity has a non-canonical stress tensor,  but which does obey the trace equation  $\Tr T = \frac{\lambda }{2\pi} \det T$ that defines the $\TTb$ flow.   However, we show that adding an improvement term to the stress tensor that preserves the trace equation does not change the deformed action, possibly after performing a field redefinition, which explains why the Nambu-Goto action also arises in the cutoff gravity context.

Assuming the standard Gibbons-Hawking-York plus cosmological constant boundary term, the  boundary stress tensor  at finite cutoff obeys the trace equation $\Tr T = \frac{\lambda }{2\pi} \det T$, where $\lambda$ is proportional to the radial cutoff location $r_c$.   This equation can be used to fix the explicit form of the non-canonical stress tensor that appears in the Nambu-Goto action.   In particular, starting from the form of the stress tensor at $\lambda =0$, including its improvement term, one can work out the stress tensor order by order in $\lambda$ by imposing the trace equation.  We carry this out, showing how to reproduce the results obtained in \cite{Kraus:2021cwf,Ebert:2022cle}.   This stress tensor is nonlocal, in the sense that each successive order in $\lambda$ involves terms with more derivatives (we do not have a closed form expression to all orders).     This nonlocal property aligns with previous results, such as the ``Dirac string'' picture developed in \cite{Cardy:2019qao}.

Having understood the emergence of the Nambu-Goto action from gravity at the classical level, we turn to a discussion of its quantization.\footnote{  The general problem of quantizing the Nambu-Goto action in general target space dimension $D$ has received much attention in the context of ``effective strings'', e.g. \cite{Aharony:2013ipa,Dubovsky:2012sh,Dubovsky:2012wk,Dubovsky:2015zey,Dubovsky:2017cnj,Dubovsky:2018bmo}, and the works of Dubovsky et. al. developed the connection to the $\TTb$ deformation.}    One of the key facts about $\TTb$ deformed CFT is that the energy spectrum of the deformed theory on the circle is related in a simple way to that of the undeformed CFT.  In particular,  $E_n ~\rt ~ (1-\sqrt{1- 2 \lambda E_n + \lambda^2 P_n^2})/\lambda$ where $E_n$ and $P_n$ are the energy and momentum of the state $n$ respectively.  One approach to obtain this, which we review, is to view the Nambu-Goto action as the gauge fixed version of the coordinate invariant Nambu-Goto action for a string in a flat $d=3$ target spacetime, which can  alternatively be expressed in light cone gauge where the action is quadratic and hence easily quantized.   At the classical level this is a canonical transformation \cite{Jorjadze:2020ili}; quantum mechanically the equivalence is somewhat murkier given that we are not in the critical dimension $d=26$.\footnote{The usual Lorentz anomaly is trivially absent in $d=3$, so the procedure is perhaps justified in this case as well.}  Regardless, this route does lead to the correct spectrum.  We also discuss the computation of the spectrum by applying standard time-independent quantum mechanical perturbation theory.   Without additional input, this leads to divergent sums over intermediate states, rendering the spectrum ambiguous, as one would expect for a non-renormalizable theory.   The additional ingredient needed to fix the spectrum is the factorized form of the trace equation $ \langle  T_{z\zb} \rangle = -\frac{\lambda }{2\pi}( \langle T_{zz} \rangle \langle T_{\zb\zb}\rangle -  \langle T_{z\zb} \rangle \langle T_{z\zb}\rangle) $, where the expectation value is taken in an energy/momentum eigenstate.  Of course, this is no surprise since this is the key relation used in \cite{Smirnov:2016lqw} to obtain the spectrum.  

We finally turn to a discussion of boundary  correlation functions, both of the elementary fields and of the stress tensor.     The Nambu-Goto action is non-renormalizable, yet viewed as a $\TTb$ deformed CFT its energy spectrum is unambiguous.  To what extent does this feature carry over to correlation functions?  Here the gravity picture provides a useful guide.  Pure 3d gravity is renormalizable: using the equations of motion, all divergences can be absorbed into the cosmological constant term \cite{Witten:2007kt}.  This leads to the expectation that on-shell  correlators of elementary fields should be unambiguous.  Indeed, existing perturbative  computations of correlators are rendered finite by counterterms that vanish on-shell.  The S-matrix is therefore unambiguous, which is a well-known feature of $\TTb$ deformed theories.\cite{Smirnov:2016lqw,Cavaglia:2016oda,Dubovsky:2012wk}

For the stress tensor the story is a bit different due to the inherent ambiguity in fixing the improvement terms.  For the gravity theory with a cutoff boundary nothing prevents us from adding a boundary term of the form $\int\! d^2x \sqrt{h}  R^{(2)}(h)Z(K)$ where $R^{(2)}(h)$ is the Ricci curvature  of the boundary and $Z(K)$  some arbitrary function of the extrinsic curvature.\footnote{This term vanishes in the asymptotically AdS limit, which is why it is usually not considered.}   The effect of this term is to add an improvement term to the stress tensor. In a general renormalizable QFT the stress tensor is renormalized by such improvement terms \cite{Brown:1979pq}, and the same is true here.   Although the stress tensor is ambiguous this does not affect the usefulness of the equation  $ \langle  T_{z\zb} \rangle = -\frac{\lambda }{2\pi}( \langle T_{zz} \rangle \langle T_{\zb\zb}\rangle -  \langle T_{z\zb} \rangle \langle T_{z\zb}\rangle $ in determining  the energy spectrum because the improvement terms have vanishing expectation value.   One might ask whether the form of the quantum stress tensor could be fixed by imposing the operator equation $\Tr T = \frac{\lambda }{2\pi} \det T$, but we show that this is not the case.  The map to light-cone gauge also does not fix these ambiguities; since it involves a state-dependent coordinate transformation, it does not map correlation functions of local operators in light-cone gauge to static gauge correlation functions of local operators.  We conclude that although off-shell correlators of the elementary fields and the stress tensor can be computed in perturbation theory their definition is subject to ambiguities.  This might be taken as an indication that these are the wrong observables to be looking at if the goal is to define the theory at a non-perturbative level.  

The rest of this paper is organized as follows.  In section \rfs{sec2} we review the boundary gravity theory in the asymptotically AdS case, and then discuss its quantization and Hilbert space in detail.  In section \rfs{sec3} we discuss the construction of the boundary stress tensor, emphasizing how its form can be fixed order by order by imposing the trace equation.  The transformation to light cone gauge is reviewed in section \rfs{sec4}, and in section \rfs{sec5} we comment on the use of ordinary perturbation theory.  Correlation functions are discussed in section \rfs{sec6} and two appendices  contain some technical details.

\section{Boundary theory at zero cutoff}
\label{sec2} 

In this section we review the boundary theory of pure three-dimensional gravity without a cutoff. The boundary theory reduces to a free scalar field with an improved stress tensor of linear dilaton (or background charge) type,  and a gauge symmetry. This will be particularly useful for later sections where we study the $T\bar{T}$ deformation of this theory. 

\subsection{Gravity solutions} 

We start with the three-dimensional Euclidean action 
\begin{equation}\label{Sdef} 
    S = -\frac{1}{16 \pi G} \int \dd^3 x \sqrt{g} \left(R+\frac{2}{\ell^2} \right) + S_{\rm bndy}~,
\end{equation}
where $S_{\rm bndy}$ is a boundary contribution that ensures a well-defined variational principle . It is well-known that in three dimensions there are no propagating gravitons in the bulk and the theory has only boundary degrees of freedom. We consider solutions  to Einstein equations with  asymptotic AdS$_3$ boundary conditions \cite{Banados:1998gg}, 
\eq{wa}{ ds^2 = \frac{dr^2 }{4r^2} +\frac{1}{r} \left[ dz-\frac{6}{c_0}r T_{\zb\zb}(\zb)d\zb \right] \left[ d\zb-\frac{6}{c_0}r T_{zz}(z)dz \right]~,}
where 
\eq{qqb}{ c_0 = \frac{3\ell}{2G}~,}
is the Brown-Henneaux central charge \cite{Brown:1986nw}.  We are working in Euclidean signature with $z = x+it$, and $x \cong x+ 2\pi$ i.e., the boundary has the topology of a cylinder.
The free functions $(T_{zz}(z),T_{\zb\zb}(\zb))$  can be identified as components of the boundary stress tensor. The mass and  angular momentum are 
\eq{waa}{ M &= \int_0^{2\pi} \frac{dx}{2\pi} T_{tt} = - \int_0^{2\pi} \frac{dx}{2\pi} (T_{zz}+ T_{\zb\zb} )~, \cr
J & =  \int_0^{2\pi} \frac{dx}{2\pi} iT_{xt} = -\int_0^{2\pi} \frac{dx}{2\pi} (T_{zz}- T_{\zb\zb} )~.    }
The stationary and rotationally symmetric solutions are written in terms of two constants $(a,\bar{a})$ as
\eq{wab}{ T_{zz} = \frac{c_0 a }{24}~,\quad  T_{\zb\zb} = \frac{c_0 \bar{a} }{24}~, } 
and carry 
\begin{align}
\label{eq:MJ2fbf}
    M = -\frac{c_0}{24} (a + \bar a), \quad J= -\frac{c_0}{24} (a- \bar a )~.
\end{align}
Global AdS$_3$ corresponds to $a=\bar{a}=1$.  Upon continuation to Lorentzian signature there are two braches of ``healthy'' solutions: rotating BTZ black holes and conical defects; see e.g. \cite{Briceno:2021dpi}.   Rotating BTZ black holes are given by $a, \bar{a}<0$, with the extremal case $M=|J|$ occurring when one of $(a,\bar{a})$ vanishes.  Conical defects have $0< a=\bar{a} < 1$.   Taking $a=\bar{a}>1$ yields ``conical excess solutions'' whose energy lie below global AdS$_3$ and  correspond to non-unitary representations of Virasoro.  Taking $a\neq \bar{a}$ with one or both $(a,\bar{a}) $ being positive  can be seen to yield naked closed timelike curves or singular horizons.
More general solutions that are ``dressed with boundary gravitons'' are obtained by taking 
\eq{wb}{ T_{zz} = \frac{c_0}{12} \left( \frac{a}{2} F'(z)^2 + \{F(z),z\}\right)~,\quad  T_{\zb\zb} = \frac{c_0}{12} \left( \frac{\bar{a}}{2} \Fb'(\zb)^2 + \{\Fb(\zb),\zb\}\right)~,}
where the  Schwarzian derivative is 
\eq{wc}{ \{F(z),z\} = \frac{F'''}{F'}-\frac{3 F''^2 }{2F'^2}~. }
The functions $(F,\Fb)$ are each elements of diff$(S^1)$, i.e., maps from the circle to itself.  They are correspondingly monotonic and obey the winding conditions
\begin{equation}
    F(x+2\pi,t) = F(x,t) + 2\pi, \quad \bar{F}(x+2\pi,t) = \bar{F}(x,t) + 2\pi~.
\end{equation}
Solutions with the same $(a,\bar{a})$ lie on a common diff orbit.  These orbits are symplectic manifolds, and the phase space action governing them is the so-called   
Alekseev-Shatashvili action 
\begin{equation}
    S= -\frac{c_0}{24 \pi} \int\dd^2 x \left[  a F' \pd_{\bar{z}} F - \left(\frac{1}{F'} \right)^{''} \pd_{\bar{z}} F + \bar{a} \bar{F}' \pd_{z} \bar{F} - \left(\frac{1}{\bar{F}'} \right)^{''} \pd_{z} \bar{F}\right]~. \label{AS action}
\end{equation}
This action has  been obtained from gravity in both  the  Chern-Simons  and metric formulations. 
In this action $(F,\Fb)=(F(x,t),\Fb(x,t))$ are each arbitrary functions of $(x,t)$, subject to the winding and monotonicity constraints.   
Primes denote $x$-derivatives and 
\begin{equation}
    \pd_{z} = \frac{1}{2} (\pd_x - i \pd_t), \quad \pd_{\bar{z}} = \frac{1}{2} (\pd_x + i \pd_t)~.
\end{equation}
In addition, \eqref{AS action} has the gauge redundancy\footnote{For the special value $a=1$ the gauge redundancy is enhanced to $PSL(2,R)\times PSL(2,R)$ \cite{Cotler:2018zff}.\label{fn:PSL}}
\begin{equation}
    F(x,t) \sim F(x,t) + \e(t), \quad \bar{F}(x,t) \sim \bar{F}(x,t) + \bar{\e}(t)~.
\end{equation}
These gauge redundancies  arise from the fact that they are invariances of the stress tensor components.  
One can easily check that the Lagrangian transforms by a total spatial derivative.

We now wish to quantize this theory.   
In the following we are going to treat the cases of positive and negative $(a,\bar{a})$ separately. In both cases the Hilbert space is a single unitary representation of the  Vir $\times $ Vir algebra, with lowest weight states determined by $(a,\bar{a})$. We refrain from considering cases with opposite signs for $(a,\bar{a})$, since the corresponding gravity solutions are pathological. 

\subsection{BTZ branch:~\texorpdfstring{$a$ and $\bar a$}{a and ā} negative}
We write 
\eq{qqa}{  a = -b^2~,\quad \bar a = -\bar b^2~. }
It was noted in \cite{Alekseev:1988ce,Alekseev:1990mp} that the field redefinition 
\begin{equation}
    \left( e^{b F} \right)' = b \;e^{f}, \quad \left( e^{\bar b \bar{F}} \right)' = \bar b\;e^{ \bar{f}} ~, \label{field reder}
\end{equation}
brings the action \eqref{AS action} to the simple form 
\begin{equation}
    S= \frac{c_0}{24 \pi }\int\dd^2 x \left( f' \pd_{\bar{z}} f + \bar{f}' \pd_{z} \bar{f}\right) ~. \label{Free ffb}
\end{equation}
This is the action of a free scalar field in the first-order formalism. From \eqref{field reder} we can read of the periodicity of the new variables 
\begin{equation}
    f(x+2\pi,t) = f(x,t) + 2\pi b , \quad \bar{f}(x+2\pi,t) = \bar{f}(x,t) + 2\pi \bar b~, \label{periodicity}
\end{equation}
while the gauge redundancy is 
\begin{equation}
    f(x,t) \sim f(x,t) + b\e(t), \quad \bar{f}(x,t) \sim \bar{f}(x,t) +\bar b \bar{\e}(t)~. \label{gauge ffb}
\end{equation}
The final element needed to describe the classical boundary theory is the boundary stress tensor. In \cite{Cotler:2018zff} it was shown to be
\begin{equation}
    T= \frac{c_0}{12} \left(-\frac{1}{2} f^{'2} + f'' \right), \quad \bar{T}= \frac{c_0}{12} \left(-\frac{1}{2} \bar{f}^{'2} +\bar{f}'' \right) ~. \label{T and Tb}
\end{equation}
To make the connection with the free scalar theory more explicit let us define 
\begin{equation}
\label{eq:f2phi}
    \f = \sqrt{\frac{c_0}{48\pi}}  (f +\bar{f}), \quad \Pi = \sqrt{\frac{c_0}{48\pi}} (f' - \bar{f}')~.
\end{equation}
The action \eqref{Free ffb} then takes the form 
\begin{equation}
\label{eq:Sphi}
    S= \int \dd^2 x \left(i \dot{\f} \Pi + \frac12 \f^{'2}+ \frac12 \Pi^2  \right)~.
\end{equation}
Naively integrating out $\Pi$ leads to the usual free scalar Lagrangian. However, this is not quite correct because \eqref{periodicity} leads to the constraint 
\begin{equation}
    \int_0 ^{2\pi} \dd x \; \Pi = \sqrt{\frac{\pi c_0}{12}} (b - \bar b)~, \label{constraint}
\end{equation}
which must be taken care of before employing the Euler-Lagrange equations. This is the first-class constraint that generates the gauge symmetry \eqref{gauge ffb}:
\begin{equation}
    \f(x,t) \rightarrow \f(x,t) + \ee(t), \quad \Pi(x,t) \rightarrow \Pi(x,t)~, \label{gauge phi}
\end{equation}
where $\ee(t) =\sqrt{\frac{c_0}{48\pi}}( b \e(t) + \bar b \bar{\e}(t))$. The transformation of the Lagrangian \eqref{eq:Sphi} is a total time derivative by virtue of the constraint \eqref{constraint}. To proceed, we first need to  pick a gauge, solve the constraint, and then use the Euler-Lagrange equations. The net effect will be that the zero modes of $\f$ and $\Pi$ are fixed. Let us explain this further. We start by expanding the fields into modes
\begin{align}
\f &= \f_0(t) + (b + \bar b) \sqrt{\frac{c_0 }{48 \pi }} x +\frac{i}{2 \sqrt{ \p}} \sum_{n\neq 0} \frac{\phi_n (t)}{n} e^{i n x} ~, \\
\Pi &= \p_0(t) + \sum_{n\neq 0} \p_n (t)e^{i n x}~,
\end{align}
where we used \eqref{periodicity} to fix the winding around the spatial circle. The constraint \eqref{constraint} fixes $\p_0 = (b - \bar b) \sqrt{c_0 / 48\pi}$. The gauge symmetry can be used to choose $\dot \f_0 = i \pi_0$, i.e., the on-shell value it would have in the absence of the constraint \eqref{constraint}. We can use the Euler-Lagrange equations for the remaining non-zero modes. Hence, we see that we end up with a free scalar field with fixed winding and zero-modes. The equations of motion then lead to the free field expansion
\begin{align}
    \f & =  \sqrt{\frac{c_0}{ 48 \pi}} [ (b + \bar b) x + i (b - \bar b) t] + \frac{i}{2 \sqrt{\p}} \sum_{n\neq 0} \frac{ \a_n}{n} e^{i n (x+it)} -\frac{i}{2\sqrt{\p}} \sum_{n\neq 0} \frac{ \bar \a_n}{n} e^{-i n (x-it)} ~,  \\
    \Pi & = (b - \bar b) \sqrt{\frac{c_0}{48\pi}} -\frac{1}{2\sqrt{\p}}\sum_{n\neq 0} \a_n e^{i n (x+it)} +\frac{1}{2\sqrt{\p}} \sum_{n\neq 0} \bar \a_n e^{-i n (x-it)} ~. \label{free field exp}
\end{align}
One can use \eqref{T and Tb} to write down a gauge-invariant expression for the stress  tensor in terms of spatial derivatives of $\phi$ and $\Pi$. However, the gauge choice we made above $(\dot \f_0 = i \pi_0)$ allows as to use the usual form
\begin{equation}
\label{eq:freeBosonT}
    -\frac{1}{2\pi} T_{\m\n}= \pd_\m \f \pd_\n \f - \frac12 \d_{\m\n} (\pd \f )^2 - \sqrt{\frac{ c_0}{12 \p}} (\pd_\m \pd_\n - \d_{\m\n} \pd^2) \f~.
\end{equation}
keeping Lorentz symmetry manifest. Thus, we conclude that the boundary theory includes also an improvement term for the stress tensor. As  is well-known this affects the representations of the boundary Virasoro algebra. The spectrum of the theory is the same with or without the improvement terms but it organizes itself differently into Virasoro modules.

The modes of the stress tensor in \eqref{T and Tb} are
\begin{equation}
    L_n=   -\oint \frac{\dd x}{2\pi} \; e^{-i n (x + i t)} T_{zz}=  \frac{1}{2}\sum_{k \neq 0} \a_{n-k}\a_{k}+(in-b) \sqrt{\frac{c_0}{12}}\a_{n} + \frac{c_0 b^2}{24} \d_{n,0}~, \label{Ln}
\end{equation}
Conjugation acts as 
\begin{equation}
\label{eq:aDag}
    \a_n^\dagger = \a_{-n}^\dagger \Rightarrow L_n^\dagger = L_{-n}^\dagger~.
\end{equation}
So far the discussion was classical. To quantize the theory we need to impose the usual commutation relations
  
\begin{equation}
    \left[ \phi(t,x), \Pi(t,x') \right] = - i \d(x-x'), \quad  \left[ \Pi(t,x), \Pi(t,x') \right]=0= \left[ \phi(t,x), \phi(t,x') \right]~, \label{quant cond}
\end{equation}
which lead to 
\begin{equation}
    \left[ \a_n, \a_m \right] = n \d_{n+m}=\left[ \bar \a_n, \bar \a_m \right], \quad  \left[ \a_n, \bar \a_m \right] =0 ~. \label{comutators}
\end{equation}
We also choose normal ordering of the operators $\a_{-k} \a_k$ for $L_0$ in \eqref{Ln}.
Using the above, one can show that the $L_n$'s obey  the algebra
\begin{equation}
    \left[ L_n, L_m\right] =(n-m) L_{n+m} + \frac{1}{12}(n^3-n) \d_{n+m}+ \frac{c_0}{12}n^3~,
\end{equation}
while the operators 
\begin{equation}
    L'_n  = L_n +\frac{c_0}{24} \d_{n,0} \label{L'n}~,
\end{equation}
satisfy the usual Virasoro algebra with central charge $c=c_0+1$,
\begin{equation}
    \left[ L'_n, L'_m\right] =(n-m) L'_{n+m} + \frac{c_0+1}{12}(n^3-n) \d_{n+m}~.  \label{Virasoro}
\end{equation}
We see that the improvement term in \eqref{eq:freeBosonT} shifts the central charge by $c_0$. To analyze the Hilbert space, note that all the $\a_n$ modes can be written as combinations of the $L'_n$ using \eqref{Ln} recursively. Hence the theory has a single primary state, namely the ``vacuum'' defined by
\begin{equation}
    \a_n \ket{0} = 0, \quad n>0.
\end{equation}
The $L_n$ operators do not obey the usual Virasoro algebra. Only $L'_n$ do, and we conclude that the Hilbert space of the theory consists of a single Virasoro module with respect to the $L'_n$'s. The primary state satisfies
\begin{equation}
     L'_0 \ket{0} = \frac{c_0(1+ b^2)}{24}  \ket{0}~, \quad  \bar{L}'_0 \ket{0} = \frac{c_0(1+ \bar b^2)}{24}  \ket{0}~.
\end{equation}
Notice that this state is not annihilated by $L_{-1}=L'_{-1}$. From \eqref{L'n} we see that the energy of the primary state is 
\begin{equation}
    (L_0 + \bar{L}_0)  \ket{0} =\frac{c_0 (b^2 + \bar b^2)}{24}  \ket{0}.
\end{equation}

\subsection{\texorpdfstring{$a$ and $\bar{a}$ positive}{a or ā positive}}
In this case
\eq{qqd}{ a = b^2~,\quad \bar a = \bar b^2~,}
and  we modify \eqref{field reder} to
\begin{equation}
    \left( e^{i b F} \right)' = i b \;e^{f}, \quad \left( e^{- i \bar b \bar{F}} \right)' = -i \bar b\;e^{\bar{f}} ~. \label{field redef 2}
\end{equation}
As can be seen after Wick rotating to Minkowski space, both $F$ and $\bar F$ are Hermitian. From the above definition it then follows that $f$ and $\bar f$ obey complicated reality properties which we analyze below. 

One can check that  this again  leads to the free scalar action  \eqref{Free ffb}. However, in this case the periodicity of the new fields is 
\begin{equation}
    f(x+2\pi,t) = f(x,t) + 2\pi i b , \quad \bar{f}(x+2\pi,t) = \bar{f}(x,t) - 2\pi i \bar b~, \label{periodicity 2}
\end{equation}
while the gauge redundancy is 
\begin{equation}
    f(x,t) \sim f(x,t) +i b\e(t), \quad \bar{f}(x,t) \sim \bar{f}(x,t) -i \bar b \bar{\e}(t)~. \label{gauge ffb 2}
\end{equation}
As a result, $\phi$ has imaginary winding
\begin{align}
    \phi(x + 2\pi, t) \to \phi(x, t) + i \sqrt{\frac{c_0 \pi}{12}} (b - \bar b)
    \ ,
\end{align}
Since $f$ and $\bar f$ are not hermitian, $\phi$ also does not have this property, hence the appearance of the imaginary winding. However, this does not imply the doubling of its modes as in the case of a complex $\phi$. On the contrary, $\phi$ has more complicated reality properties and, as we will see later, this translates to an unconventional conjugation for its modes.

In this case the momentum $\Pi$ obeys the constraint
\begin{equation}
    \int_0 ^{2\pi} dx \; \Pi = i (b + \bar b) \sqrt{\frac{c_0 \pi}{12}}~,
\end{equation} 
while the gauge symmetry acts as before \eqref{gauge phi} where now $\ee(t) = \sqrt{\frac{c_0}{48\pi}} ib (\e(t)-\bar{\e}(t))$. We can proceed as before and write the free field expansion
\begin{align}\label{pex}
    \f & = \sqrt{\frac{c_0}{48\pi}} \left[ i (b - \bar b) x - (b + \bar b) t \right] + \frac{i}{2 \sqrt{\p}} \sum_{n\neq 0} \frac{ \a_n}{n} e^{i n (x+it)} -\frac{i}{2\sqrt{\p}} \sum_{n\neq 0} \frac{ \bar \a_n}{n} e^{-i n (x-it)} ~,  \\
    \Pi & =i (b + \bar b) \sqrt{\frac{c_0}{48 \pi }}  -\frac{1}{2\sqrt{\p}}\sum_{n\neq 0} \a_n e^{i n (x+it)} +\frac{1}{2\sqrt{\p}} \sum_{n\neq 0} \bar \a_n e^{-i n (x-it)} ~. \label{free field exp 2}
\end{align}
In this case the modes of the stress tensor are
\begin{equation}
    L_n=  \frac{1}{2}\sum_{k \neq 0} \a_{n-k}\a_{k}+i (n-b) \sqrt{\frac{c_0}{12}}\a_{n} - \frac{c_0 b^2}{24} \d_{n,0}~. \label{Ln 2}
\end{equation}
As before, to quantize the theory we impose the commutation relations \eqref{quant cond} which again lead to \eqref{comutators}. However, conjugation in this case needs to be modified. To preserve the action of conjugation on the $L_n$'s 
\begin{equation}
    L_n^\dagger = L_{-n}~,
\end{equation}
we need to define
\begin{equation}
    \a_n^\dagger =  \frac{n+b}{n-b} \a_{-n} - i\frac{nb}{n-b} \sqrt{\frac{12}{c_0}}\sum_{k\neq 0} \frac{1}{(b+k)(b-k-n)} \a_{-n-k}\a_{k}  + \dots~, \label{phi dagger}
\end{equation}
where the dots denotes terms with more $\a_n$'s. Higher order terms are accompanied with higher powers of $1/c_0$. For large $c_0$ one can check that for the first few orders, \eqref{phi dagger} is consistent with
\begin{equation}
    \left[ \a_n, \a_m \right] = n \d_{n+m} \Rightarrow \left[ \a_n^\dagger, \a_m ^\dagger\right] = -n \d_{n+m}~.
\end{equation}
Equation \eqref{phi dagger} is valid only for $0 < b < 1$, but not for global AdS with $b = \bar b = 1$ in which case there is additional gauge symmetry.\textsuperscript{\ref{fn:PSL}} It is surprising that one is forced to consider this peculiar definition of conjugation. It would be interesting to understand its geometric interpretation from the bulk 3d gravity point of view.

Let's now consider the norm of the state $\a_{-n} \ket{0}$ for $n>0$. Using the above expression for $\phi^\dagger_n$ one can show that only the first term contributes 
\begin{equation}
    || \a_{-n} \ket{0} ||^2 =  \bra{0} \a_{-n}^\dagger \a_{-n} \ket{0} || = \frac{ n(n-b)}{n+b}~.
\end{equation}
We see that the spectrum is unitary only for $ -1 \leq b \leq 1$. At the endpoints of this interval null states appear. Combining with the results of the previous section we see that the spectrum is unitary only for $a<1$.\footnote{Solutions in gravity with $a,\bar{a} >1$ correspond to ``conical excesses'', and are indeed associated with non-unitary representations \cite{Castro:2011iw,Raeymaekers:2020gtz}.}  Note the slightly curious fact that we obtain unitary representations of Virasoro in this range  for $a\neq \bar{a}$ even though the corresponding gravity solutions have naked closed timelike curves.  The only novel aspect of this parameter regime is the fact that $\phi$ in \rf{pex} is not Hermitian. 

As in the case with negative $a$, one can similarly define \eqref{L'n} which obey the usual Virasoro algebra. We then conclude that the Hilbert space of the theory contains a single Virasoro module with

\begin{equation}
     L'_0 \ket{0} =  \frac{c_0(1-b^2)}{24}  \ket{0} \ , \quad \bar{L}'_0 \ket{0} = \frac{c_0(1-\bar b^2)}{24}  \ket{0}.
\end{equation}
For $b, \bar b<1$ this is a unitary representation whose primary state has energy
\begin{equation}
    (L_0 + \bar{L}_0)  \ket{0} =-\frac{c_0 (b^2 + \bar b^2)}{24}  \ket{0}.
\end{equation}

\subsection{Thermal partition function}

We end this section with a few comments. In the previous sections we showed that the bulk theory with the boundary condition \eqref{wa}, where $T_{zz}$ and $T_{\zb\zb}$ given by \eqref{wab}, leads to a single Virasoro primary on the boundary that satisfies 

\begin{equation}
    L'_0\ket{0} = \frac{c_0(1-a)}{24}\ket{0}, \quad \bar{L}'_0\ket{0} = \frac{ c_0(1-\bar{a})}{24}\ket{0}~.
\end{equation}

In our analysis the boundary was a cylinder. Let's now compactify also the time direction   and consider the theory  on the torus. If we define the partition function as 
\begin{equation}
    Z = (q\bar{q})^{-\frac{1}{24}}\Tr q^{L_0} \bar{q}^{\bar{L}_0} =(q\bar{q})^{-\frac{c_0+1}{24}}\Tr q^{L'_0} \bar{q}^{\bar{L}'_0} ~,
\end{equation}
we conclude that it is equal to

\begin{equation}
    Z =\chi_\frac{c_0 (1-a)}{24} (q)\chi_\frac{c_0 (1-\bar{a})}{24} (\bar{q})~,
    \label{Z}
\end{equation}
where $\chi_h(q)$ is the character of a module of scaling dimension $h$. 
This has been obtained in the literature before, either by computing the partition function of the Alekseev-Shatashvili  theory in the path integral formulation \cite{Cotler:2018zff} or by computing it directly in the bulk \cite{Maloney:2007ud,Giombi:2008vd}.

Since the above expression \eqref{Z}  contains just a single Virasoro character it cannot be the partition function of a modular invariant CFT.  From the gravity point of view we would expect that black holes of different masses and angular momenta should be part of a common Hilbert space and so should be included in the sum over states.  However, gravity provides little guidance as to what spectrum of masses to include. The spectrum should be discrete in order to obtain a finite  partition sum, and should also be in approximate agreement with the Bekenstein-Hawking entropy formula.  The alternative sum over Euclidean geometries can produce a modular invariant result but suffers from other pathologies  \cite{Maloney:2007ud,Benjamin:2019stq}. These facts have led to the suggestion that the holographic dual of 3d gravity in AdS$_3$ is actually an ensemble of CFTs  \cite{Afkhami-Jeddi:2020ezh,Maloney:2020nni,Chandra:2022bqq}.

\section{The stress tensor in cutoff gravity and \texorpdfstring{$\TTb$}{TTbar}-deformed CFT}
\label{sec3}

\subsection{\texorpdfstring{$\TTb$}{TTbar} deformation and stress tensor improvement} 

At the classical level the $\TTb$ deformation corresponds to finding an action $S_\lambda(\phi)$ that obeys the flow equation 
\eq{ua}{  \frac{dS_\lambda }{d\lambda}= -\frac{1}{8\pi^2} \int\! d^2x \det T~,}
where the factor is $\frac{1}{8\pi^2 } $ is convention dependent.    The stress tensor $T_{\mu\nu}$ appearing on the right hand side should be conserved with respect to the equations of motion of $S_\lambda$.  However, given one conserved stress tensor $T_{\mu\nu}$ one can always write down another conserved stress tensor by adding an improvement term,\footnote{Although this is not the most general improvement term for a generic QFT, it is for the case under considerations which involves only a single scalar field.}
\eq{ub}{ \tilde{T}_{\mu\nu} =  T_{\mu\nu}  + (\pd_\mu \pd_\nu -\eta_{\mu\nu} \pd^2)Y(\phi)~.}
Different choices for the stress tensor might therefore seem to lead, via \rf{ua}, to different expressions for the deformed action.  However, we now show that these actions differ at most by a field redefinition.

We first note  that in a deformed CFT, $\lambda$ is the only dimensionful scale, and since $\Tr T$ is the generator of scale transformations we have that $\lambda \frac{dS_\lambda }{d\lambda} = -\frac{1}{4\pi} \int \! d^2x \Tr T $, which implies that $\frac{\lambda}{2\pi} \det T =   \Tr T $, up to a possible total derivative. On the other hand, assume that we define the deformed action in \eqref{ua} using the improved stress tensor $\tilde{T}_{\m\n}$. For the reason we just explained, $\tilde{T}_{\m\n}$ will also obey a trace equation $\frac{\lambda}{2\pi} \det \tilde{T} =  \Tr \tilde{T} $ up to a total derivative. From \rf{ub} we have that $\Tr \tilde{T} = \Tr T  -  \pd^2 Y$  from which it follows that $\det T = \det \tilde{T} $ up to a total derivative. Therefore $S_\lambda$ also obeys equation \rf{ua} but with $T_{\mu\nu}$ replaced by $\tilde{T}_{\mu\nu}$.

This observation clarifies one of the main results of \cite{Ebert:2022cle}.  It is well-known that the deformed action for a free scalar with canonical stress tensor is the Nambu-Goto action.  On the other hand, the undeformed action in \cite{Ebert:2022cle} is a free scalar but the stress tensor is not canonical, rather it has a linear dilaton improvement term.  The argument above explains why the deformed action starting from this non-canonical stress tensor turns out to be the same Nambu-Goto action as in the canonical case.

Actually, there is one important point in this argument that we glossed over. 
The trace equation $\frac{\lambda}{2\pi}  \det T = \Tr T $ is satisfied only on-shell and therefore the statement $\det T = \det \tilde{T} $ holds only after we use the equations of motion.
If one uses two stress tensors that differ by terms that vanish on-shell then the corresponding deformed actions may also differ by such terms.  However, on general grounds one knows that two actions that differ in this way are related by a field redefinition; e.g \cite{Arzt:1993gz}.    So the general statement is that the ambiguity regarding which stress tensor to use in  \rf{ua} does not translate into any ambiguity in the action, assuming one uses the freedom to make field redefinitions. 
Indeed, in \cite{Ebert:2022cle} a complicated field redefinition, not known in closed form, was needed in order to bring the action to the Nambu-Goto form.

\subsection{Nambu-Goto action from the trace equation} 

In classical cutoff gravity with the standard boundary term (GHY plus boundary cosmological term) the stress tensor obeys the trace equation 
\begin{align}
\label{eq:Tr}
T^\mu_\mu = \frac{\l}{2\pi} \det T^\mu_{~\nu}~,
\end{align}
We now quickly review how this deforms a free scalar field into the Nambu-Goto action.  We will use the canonical stress tensor, deferring the discussion of the improvement term to the next section.  From the argument of the previous section we know that we will obtain the same action in either case.   Since the undeformed action and the canonical stress tensor contain only first derivatives of the scalar field, \eqref{eq:Tr} will not generate any dependence on higher derivatives. To preserve this feature, we do not add a total derivative term to \eqref{eq:Tr}.  

To follow the above strategy it is more convenient to write the action in terms of $\phi$ and its derivatives than in terms of $\phi$ and its canonical momentum $\Pi$. In this approach Lorentz symmetry is manifest and we can easily get a closed form expression for the action to all orders in the deformation parameter $\l$.  A subtlety in this formulation is that the Lagrangian is not gauge invariant.  However, once we convert back to canonical variables the gauge symmetry is restored. The strategy we follow is to impose that the Lagrangian and the stress tensor in terms of  $\phi$  and its derivatives are invariant under
\begin{equation}
    \phi \rt \phi + \e~, \label{constant gauge}
\end{equation}
for constant $\e$. Equivalently we require that they do not depend on $\phi$ itself but only on its derivatives.  After obtaining the deformed action, we replace time derivatives of $\phi$ with $\Pi$ to find expressions that are gauge invariant under \eqref{gauge phi}. It is possible to directly repeat the derivations in this section using $\phi$ and $\Pi$ but it is not very illuminating. We have included it in  appendix  \ref{appA}.

From dimensional analysis and Poincaré invariance, the action must take the form
\begin{align}
\label{eq:S2F}
    S = \frac1\l \int d^2 x \, F(s)~,\quad 
    s \equiv \l \, \pd_\mu \phi \, \pd^\mu \phi
    \ .
\end{align}
The canonical stress tensor defined as

\begin{equation}
    T^\text{can}_{\m\n} = 2\pi\left(\d_{\m\n}L-\frac{\pd L}{\pd (\pd^\m \phi)} \pd_\n \phi\right)~, \label{can T}
\end{equation}
is then equal to 
\begin{align}
    T_{\mu\nu}^\text{can} &= \frac{2\pi}\l (\d_{\mu\nu} F - 2\l \, \pd_\mu \phi \, \pd_\nu \phi \, F')
    \ ,
\end{align}
and the trace equation \eqref{eq:Tr} reduces to a differential equation for $F$,%
\footnote{A convenient identity is
\begin{align}
    \det T &= \frac12 (\d_\mu^\rho \d_\nu^\sigma - \d_\mu^\sigma \d_\nu^\rho) T^\mu{}_\rho T^\nu{}_\sigma= \frac{1}{2}\big( (\Tr T)^2 - \Tr(T^2)\big)
    \ .
\end{align}
}
\begin{align}
    \frac{2\pi}\l (2F - 2s F') &= \frac{2\pi}\l F (F - 2s F')
    \ .
\end{align}
This ODE can be solved by integration. There are two branches and one integration constant $n$: $F = 1 \pm \sqrt{1 - n s}$. We want the action \eqref{eq:S2F} to reduce to the free boson action for $\l \to 0$, so we are forced to choose the lower sign and fix $n = 1$ to find
\begin{align}
    S &= \frac1\l \int d^2 x \left( 1 - \sqrt{1 - \l \pd_\mu \phi \, \pd^\mu \phi} \right)
    \ . \label{defomred action}
\end{align}
From the deformed action \eqref{defomred action} we can derive the canonical stress tensor
\begin{align}
    T_{\mu\nu}^\text{can} &= \frac{2\pi}\l \left[ (1 - \sqrt{1 - s}) \d_{\mu\nu} - \frac{\l \pd_\mu \phi \, \pd_\nu \phi}{\sqrt{1 - s}} \right]
    \label{deformed T}~, \\
    &= 2\pi \left[ \frac12 (\pd \phi)^2 \d_{\mu\nu} - \pd_\mu \phi \, \pd_\nu \phi \right] + \mathcal{O}(\l)~,
    \ 
\end{align}
which indeed reduces to \eqref{eq:freeBosonT} with $c_0=0$ (corresponding to no improvement term)  in the limit $\l \rt 0$.

\subsection{Computing the improved stress tensor} 

We now explain how to obtain the stress tensor that arises from cutoff gravity.   This stress tensor takes the form 
\begin{align}
    T_{\mu\nu} &= T_{\mu\nu}^\text{can} + (\pd_\mu \, \pd_\nu - \d_{\mu\nu} \pd^2) Y
    \ , \label{improved T}
\end{align}
where $T_{\mu\nu}^\text{can}$ is the canonical part \eqref{deformed T} derived from the Nambu-Goto action. More precisely, starting from the Lagrangian in terms of $\phi$ and $\Pi$, the stress tensor can be put in this form using the equations of motion. 
In the asymptotically AdS case corresponding to $\lambda=0$ we know that the improvement term is 
\begin{equation}
    Y = \beta \phi + \mathcal{O}(\l)~,\quad \beta = \sqrt{\frac{c_0 \pi }{ 3}}~. 
\end{equation}
Combining this with the fact that the stress tensor obeys the trace equation \rf{eq:Tr}, we can uniquely determine $Y$ order by order in $\lambda$, as we now show.   After using the equations of motion this procedure reproduces the stress tensor computed in \cite{Ebert:2022cle}, and thereby establishes the precise sense in which the cutoff gravity theory is the $\TTb$-deformed version of the free scalar with improved stress  tensor.  

At first order  in $\l$ it is easy to see that the only  Lorentz invariant term allowed by dimensional analysis is 
\begin{equation}
    Y = \beta \phi + \gamma \l  (\partial \phi)^2 + \mathcal{O}(\l^2)~.
\end{equation}
Note that the gauge transformation \eqref{constant gauge} does not allow polynomials $\phi^n$ with $n \neq 1$. To this order the trace of the stress tensor becomes

\begin{equation}
    \Tr T=-\beta \pd^2 \phi - \frac{\pi \l}2 (\pd \phi)^4 - \g\l\;\pd^2 ( \partial \phi )^2 + \mathcal{O}(\l^2)~, \label{trace a^1}
\end{equation}
while the right-hand side of the trace equation \eqref{eq:Tr} is equal to
\begin{equation}
    \frac{\l}{2\pi} \det T  = - \frac{\l}{4\pi}\Tr T^2+ \mathcal{O}(\l^2) = \b \l\pd_\mu \phi \, \pd_\nu \phi \, \pd^\mu \pd^\nu \phi - \frac{\pi \l}2 (\pd \phi)^4 - 
    \frac{\l \b^2}{8\pi}\;\pd^2 ( \partial \phi )^2 + \mathcal{O}(\l^2)~.\label{det a^1}
\end{equation}
We can use the equations of motion 
\begin{align}
\label{eq:phiEoM}
    \pd^2 \phi &= -\l \frac{\pd_\mu \phi \, \pd_\nu \phi \, \pd^\mu \pd^\nu \phi}{1 - \l (\pd \phi)^2} = -\l \pd_\mu \phi \, \pd_\nu \phi \, \pd^\mu \pd^\nu \phi + \mathcal{O}(\l^2)
    \ .
\end{align}
and demanding \eqref{eq:Tr} we get $\g = \frac{ \b^2}{8\pi}$. 

One can follow the same procedure and at each order in $\l$ determine the improvement term $Y$ in \eqref{improved T}. We carried out this calculation up to $\l^3$ and we show that it leads the following unique $Y$ 
\begin{align}
    Y &= \b \phi + \frac{\l \b^2}{8\pi} (\pd \phi)^2 + \frac{\l^2 \b^2}{32\pi} (\pd \phi)^4 + \frac{\l^2 \b^3}{32 \pi^2} \pd_\mu \phi \, \pd_\nu \phi \, \pd^\mu \pd^\nu \phi
    \nonumber \\
    &\quad + \frac{\l^3 \b^2}{64 \pi} (\pd \phi)^6 + \frac{5\l^3 \b^3}{192 \pi^2} \pd_\mu \phi \, \pd_\nu \phi \, \pd^\mu \pd^\nu \phi (\pd \phi)^2
    \nonumber \\
    &\quad + \frac{\l^3 \b^4}{384 \pi^3} [ \pd_\mu \phi \, \pd_\nu \phi \, \pd_\rho \phi \, \pd^\mu \pd^\nu \pd^\rho \phi + \frac32 (\pd \phi)^2 \pd_\mu \pd_\nu \phi \, \pd^\mu \pd^\nu \phi ] + \mathcal{O}(\l^4)~. \label{Y}
    \end{align}
    
We believe that the trace equation  determines $Y$ to all orders in $\l$ but unfortunately we were not able to get an answer in a closed form.  All in all we conclude that the full stress tensor is given by \eqref{improved T} with the canonical part as in \eqref{deformed T} and the improvement term as in \eqref{Y}.

To relate to results in \cite{Ebert:2022cle} we should express $\dot{\phi}$ as a function of $\phi'$ and $\Pi$ using the equations of motion following from the canonical form of the Nambu-Goto action, 
\begin{align}\label{Sform} 
    S &= \int d^2 x \left[ i \Pi \dot \phi + \frac1\l \left( 1 - \sqrt{1 - \l (\phi'^2 + \Pi^2) + \l^2 \phi'^2 \Pi^2} \right) \right]
    \ .
\end{align}
This is indeed the $T\Tb$ deformed version of the action \eqref{eq:Sphi} (see \cite{Jorjadze:2020ili}). Note that the action is gauge invariant under \eqref{gauge phi}.
We further express results in terms of $f$ and $\fb$ using \eqref{eq:f2phi}. The action in these variables is given by
\eq{zzza}{
    S &= \frac{c_0}{48 \pi }\int\dd^2 x \; \left[ i f' \dot{f} -i \bar{f}' \dot{\bar{f}} + \frac{4}{r_c} \left( 1 - \sqrt{1 - \frac12 r_c (f^{'2} + \bar{f}^{'2}) + \frac{1}{16}r_c^2(f^{'2} - \bar{f}^{'2})^2} \right) \right]    \ ,
}
where 
\eq{zza}{ r_c = \frac{ \l c_0}{12\pi}~.}
Carrying out the conversion, the canonical part of the stress tensor is given 
 \eq{ahz}{  T^{\rm can}_{z\zb} & =  -\frac{c_0}{ 96 } \left(r_c f'^2 \fb'^2 +\frac{1}{ 2} r_c^2 f'^2 \fb'^2(f'^2 +\fb'^2) +\ldots   \right)~, \cr
T^{\rm can}_{zz}  & = -\frac{c_0 }{ 24 } \left( f'^2+\frac{1}{ 2}r_c f'^2 \fb'^2 +\frac{3}{ 16} r_c^2 f'^2 \fb'^2 (f'^2+\fb'^2) + \ldots   \right)~,  \cr
T^{\rm can}_{\zb\zb} &  = -\frac{c_0}{ 24 } \left( \fb'^2+\frac{1}{ 2}r_c f'^2 \fb'^2 +\frac{3}{ 16} r_c^2 f'^2 \fb'^2 (f'^2+\fb'^2) + \ldots   \right) ~,  }
while the improvement term is
\eq{zzb}{ Y = \frac{c_0}{6} \left[ \frac{1}{2} (f+\fb)+ \frac{r_c}{4}f'\fb' +\frac{ r_c^2}{16}(f''\fb'^2+ f'^2 \fb'')+\ldots \right]~.
}
Acting with the appropriate derivatives \eqref{improved T}, the corresponding stress tensor components are
\eq{zzc}{  T_{z\zb}& = \frac{c_0}{6} \left[ -\frac{1}{4} r_c f''\fb''  + \frac{1}{ 8}r_c( f''\fb'^2+ f'^2 \fb'') - \frac{1}{ 8}r_c^2 (f'f''\fb'''+ f''' \fb' \fb'') +\ldots  \right]~, \\ 
T_{zz}& = \frac{c_0}{6} \left[\frac{1}{2}f''+  \frac{1}{ 8}r_c ( 2f'' \fb'^2 +2 f'\fb'\fb''-2 f'f'' \fb') + \frac{1}{ 8} r_c^2 (f'f''' \fb''+ f''^2 \fb'' +\frac{1}{2} f'''' \fb'^2) +\ldots  \right]~,\cr
T_{\zb\zb}& = \frac{c_0}{6} \left[\frac{1}{2}\fb''+  \frac{1}{ 8}r_c ( 2\fb'' f'^2 +2 \fb'f'f''-2 \fb'\fb'' f') + \frac{1}{ 8} r_c^2 (\fb'\fb''' f''+ \fb''^2 f'' +\frac{1}{2} \fb'''' f'^2) +\ldots  \right]~.\cr \nonumber
}
Note that when expressed in terms of $f$ and $\bar f$  the constant in front of the improvement term $c_0$ appears as an overall factor.  Indeed, since the gravity action is proportional to $\frac{1}{G}\sim c_0$ the classical stress tensor must be proportional to $c_0$.  The sum of \rf{ahz} and \rf{zzc} reproduces the result in \cite{Ebert:2022cle} obtained from the 3d bulk gravity. 

\subsection{Reduction to deformed Schwarzian quantum mechanics}

Two-dimensional JT gravity defined on a Euclidean spacetime with disk topology is described by the Schwarzian theory \cite{Maldacena:2016upp}.  The finite cutoff version of this theory was studied in \cite{Iliesiu:2020zld,Stanford:2020qhm}. In particular, in  \cite{Iliesiu:2020zld} a partial result for  the corresponding deformed Schwarzian action was written down.   We now explain how to reproduce this result from ours by dimensional reduction.  Starting from \rf{zzza} we take $(f,\overline{f})$ to depend only on $x$, and we also set $\overline{f}=f$.  Dropping the $t$ integration we get 
\eq{zzzb}{ S = \frac{c_0}{ 12 \pi r_c} \int\! dx \left( 1-\sqrt{1- r_c f'^2}\right)~.}
This action describes a finite cutoff geometry in JT gravity, as follows from the fact that the steps in the reduction are precisely those that implement the Kaluza-Klein reduction of  our 3d gravity to JT gravity \cite{Mertens:2018fds}.    On the other hand, in \cite{Iliesiu:2020zld} the cutoff JT action was presented as
\eq{zzzc}{ S =  \frac{\phi_r}{\varepsilon^2} \int\! dx \left(1- \sqrt{1+2\varepsilon^2 {\rm Sch}(z,x)}  + {\rm derivatives~of~Schwarzian} \right)~,}
where the (non-total) derivative terms were left undetermined.  To see the connection to our result  we note that the field redefinition
\eq{zzzd}{ z' =e^f  }
gives 
\eq{zzze}{ {\rm Sch}(z,x) = f''-\frac{1}{2}f'^2~.}
Using the argument in section 3.5 of \cite{Iliesiu:2020zld} that the Schwarzian is conserved, we can compare these actions for constant values of the Schwarzian. This corresponds to constant $f'$, hence to $f''=0$.  Using also the correspondence $r_c = \varepsilon^2$ and $\frac{c_0}{12\pi} = \phi_r$, we find agreement between \rf{zzzb} and \rf{zzzc}.     

The function $z(x)$ appearing in \rf{zzzc} describes directly the embedding of the cutoff boundary in AdS$_2$, however the corresponding action is only known up to terms involving derivatives of the Schwarzian, as indicated.  On the other hand, the action \rf{zzzb} is exact but the field redefinition between $f(x)$ and the embedding of the boundary is not known in closed form.  The  perturbative expansion of the field redefinition follows by reduction from the 3d field redefinition worked out (to a finite order) in \cite{Ebert:2022cle}. 

\section{Static vs light-cone gauge}
\label{sec4} 

As has been noted by various authors \cite{Caselle:2013dra,Cavaglia:2016oda,McGough:2016lol,Callebaut:2019omt,Jorjadze:2020ili}, the fact that the $T\Tb$-deformed theory of a collection of free scalar fields turns out to be the Nambu-Goto action in static gauge implies that one can solve the theory by transforming to light-cone gauge where the action is quadratic.  Of course for a general number of scalar fields this is not  justified at the quantum level due to the target space Lorentz anomaly.  However, the Lorentz anomaly is absent in two special cases: for $24$ scalar fields one has the usual cancellation of the anomaly, while for a single scalar field the anomaly is trivially absent since it must be antisymmetric in transverse target space indices.  So there is a reason to believe that the procedure is justified in the case relevant to 3d gravity, and indeed the resulting  energy  spectrum  is the correct one.

The canonical transformation between the static and light-cone gauge theories was reviewed in the $T\Tb$ context in \cite{Jorjadze:2020ili}.  The only slight difference we need to take into account is that our scalar field theory is a gauged version of the one in \cite{Jorjadze:2020ili}.  Since this difference essentially goes along for the ride, we simply state a few of the main results and add some comments.  

In this section we work in Lorentzian signature, taking our conventions as in Appendix \ref{appLor}, with   action 
\eq{ba}{ S  =  \int d^2 x \left[  \Pi \dot \phi - \frac1\l \left( 1 - \sqrt{1 - \l (\phi'^2 + \Pi^2) + \l^2 \phi'^2 \Pi^2} \right) \right]~. }
We will focus attention on the case  $(a=-b^2,\overline{a} =-\bb^2) $ for which the winding conditions are \footnote{The other sign choice leads to imaginary winding conditions and constraints, which clashes with the intepretation in terms of a string embedded in a   real target space.}
\eq{baa}{ f(x+2\pi,t) = f(x,t) +2\pi b~,\quad \fb(x+2\pi,t) = \fb(x,t)+2\pi \bb~.}
This translate into the periodicities
\eq{bab}{ \phi(x+2\pi,t) = \phi(x,t) + \sqrt{\frac{c_0}{48\pi} } 2\pi (b+\bb) ~,\quad   \Pi(x+2\pi,t)=\Pi(x,t)~, }
along with the constraint  
\eq{bb}{ \int_0^{2\pi}  \Pi(x,t) dx=\sqrt{\frac{\pi c_0}{12}}(b-\bb).}

We  view this action as a gauge fixed version of the Hamiltonian form of the Nambu-Goto action
\eq{bg}{S= \frac{1}{\l} \int\! d^2x \sqrt{-\det \pd_i X^\mu \pd_j X_\mu }  ~,}
which in Hamiltonian form reads 
\eq{bc}{S = \int\! d^2x \left[  \Pi_\mu \dot{X}^\mu - \lambda_1 C_1 -\lambda_2 C_2 \right]~, }
where the constraints are 
\eq{bd}{ C_1 = \Pi_\mu X'^\mu~,\quad C_2 = \frac{1}{2} ( \l^2 \Pi_\mu \Pi^\mu + X'_\mu X'^\mu)~.}
Here $\mu=0,1,2$ with the Minkowski metric $\eta_{\mu\nu} = (-1,1,1).$  The $X^1$ direction is taken to be compact and we consider the unit winding sector, $X^1(x+2\pi,t)= X^1(x,t)+2\pi$ .  

\subsection{Static gauge} 

Static gauge is defined by 
\eq{be}{ X^0=t~,\quad X^1 = x~.}
Solving the constraints and plugging back into \rf{bc} reproduces \rf{ba} up to a constant additive shift under the identification
\eq{bf}{X^2 = \sqrt{\l} \phi~,\quad \Pi_2 = \frac{1}{\sqrt{\l}} \Pi~.}
The constraint \rf{bb} translates to a constraint on $\Pi_2$.  In terms of the string moving in a three-dimensional target spacetime, this constraint amounts to fixing the  spacetime momentum in the $X^2$ direction, $P_2 =\int_0^{2\pi} \Pi_2(x,t) dx $.    The winding condition \rf{bab} on $\phi$ translates into 
\eq{bfa}{ X^2(x+2\pi,t) =X^2(x,t) + \sqrt{ \frac{\l c_0}{48\pi}} 2\pi (b+\bb)~. }
The string therefore winds around both cycles of a target space torus in $X^{1,2}$.

\subsection{Light-cone gauge} 

Next we consider the light-cone gauge.  Define
\eq{bgg}{ X^\pm = X^0 \pm X^1~,\quad \Pi_\pm = \frac{1}{2} (\Pi_0 \pm \Pi_1)~,}
and fix the gauge as 
\eq{bh}{ X^+(x,t) = -2\l p_- t +x~,\quad \Pi_- = p_- ~,}
where $p_-$ is constant.   The other light-cone component of the target space momentum is  $p_+ = \int_0^{2\pi} \frac{dx}{2\pi} \Pi_+$.  Integrating the constraints allows us to relate $p_\pm$ to $(\phi,\Pi)$, defined in \rf{bf}, as
\eq{bi}{ (p_+-p_-) + P_{lc}=0~,\quad 2\l H_{lc}+(1-4\l^2 p_+p_-)=0 ~,}
where
\eq{bk}{ P_{lc} = \int_0^{2\pi}  \Pi \phi' dx ~,\quad H_{lc}  = \int_0^{2\pi}  \frac{1}{2}(\Pi^2+ \phi'^2) dx ~. }
The action is 
\eq{bl}{S = \int\! d^2x \left[\Pi \dot{\phi}- \frac{1}{2}(\Pi^2+ \phi'^2)\right]  ~, }
along with a decoupled part involving the zero modes that we refrain from writing \footnote{As shown in \cite{Jorjadze:2020ili} there is also a decoupled term $\int dt p_- \dot{q}^-$. This leads to an extra label for the states, namely the eigenvalue value of $q_-$. However, in \cite{Jorjadze:2020ili} it was also shown that because of the constraints the value of $q_-$ is fixed in terms of the energy and momentum and thus no additional degeneracy is introduced. }.  The free field form of this action is of course the main virtue of the light-cone gauge.   We can view this free field theory as our undeformed theory. We have the spectra 
\eq{bla}{ H_{lc} = \frac{ c_0}{24}(b^2+\bb^2) +  N+\Nb~,\quad P_{lc} =\frac{c_0}{24}(b^2-\bb^2) +  N-\Nb~,}
where the ground state contributions  come from the momentum and winding. 

\subsection{Spectrum of static gauge Hamiltonian} 

We are interested in the energy spectrum of the static gauge theory, since the Hamiltonian of that theory is identified with that of the $T\Tb$-deformed theory.  Explicitly, 
\eq{bm}{H &= \int_0^{2\pi}  dx  \frac{  1 - \sqrt{1 - \l (\phi'^2 + \Pi^2) + \l^2 \phi'^2 \Pi^2}  }{\l}  
    \ .}
$H$ generates $t$ translations, which in static gauge is the same as $X^0$ translations.  In light-cone gauge the generator of $X^0$ translations is $-(p_++p_-)$, so the static gauge Hamiltonian expressed in terms of light-cone operators is  $H=-(p_++p_-) $.    The constraints \rf{bi} allow us write this as 
\eq{bn}{ H = \frac{1-\sqrt{1-2\l H_{lc}+\l^2 P^2_{lc} } }{\l} ~,  }
where we added the constant piece by hand.   The free field spectra of $(H_{lc},P_{lc})$ then imply that the spectrum of $H$ is 
\eq{bo}{H &=  \frac{  1 - \sqrt{1 - 2\l (\frac{ c_0}{12}(b^2+\bb^2)+  N+\Nb) + \l ^2(\frac{ c_0}{12}(b^2-\bb^2)+ N-\Nb)^2}  }{\l}  
    \ .}
This reproduces the $T\Tb$ energy spectrum.   We now make a few comments.   As shown  in \cite{Jorjadze:2020ili,Kruthoff:2020hsi} the phase space variables $(\phi,\Pi)$ in the two gauges are related by a canonical transformation.    While this establishes the equivalence of the theories at the classical level, this does not automatically extend to the quantum theory since classical canonical transformations typically have no quantum counterpart.\footnote{This fact is sometimes stated by saying that there is no natural action of symplectomorphisms on Hilbert space.}  Of course, the question of quantum equivalence of the two theories is ill-defined unless one provides an independent definition of the static gauge theory with its unwieldy square root.   The point to be emphasized here is that if we interpret the static gauge action as being a $T\Tb$ deformed theory then this fixes the spectrum of the theory and this spectrum coincides with that obtained from the light-cone theory.  

The thermal partition function can be calculated as a trace over the Hilbert space of $e^{-\beta H}$, although we are not aware of a closed form expression like \eqref{Z} in the deformed case. This is a specific example of the general analysis of torus partition functions of $T \overline T$ deformed theories \cite{Cardy:2018sdv,Dubovsky:2018bmo,Aharony:2018bad}. In particular, eq.~53 of \cite{Dubovsky:2018bmo} rewrites the finite temperature path integral in terms of a sum over the energy levels \eqref{bo} of the deformed CFT. In the other direction, one can derive a path integral expression from the canonical formalism by inserting a dense set of equal-time slices in the usual way. The resulting finite temperature path integral will, by construction, agree with the Hilbert space trace. Because the kinetic term in the action \eqref{Sform} is trivial, the path integral measure associated with the symplectic form of $f$ and $\fb$ will be flat. 

In \cite{Dubovsky:2012wk} the quantum equivalence of the static and light-cone theories was discussed in the context of the S-matrix.   The static gauge theory has a nontrivial $2 \rt 2$ S-matrix, which at first seems in conflict with the fact that the light-cone theory is free.  However this can be understood in terms of the field dependent coordinate transformation that relates the theories.  Namely, the static gauge $2\rt 2$ S-matrix is given by a time delay, and under the coordinate transformation this time delay vanishes in light-cone gauge, as it must in a free theory.  

Since correlation functions are readily computed in the light-cone theory, one might wonder whether correlators in static gauge could be computed by transforming them to light-cone gauge.  Unfortunately, the canonical transformation maps simple static gauge operators to complicated light-cone operators and vice versa, so this does not appear to be helpful.  In slightly more detail, consider a correlation function of local operators in static gauge, each evaluated at a position $x_n^s$ with superscript $s$ for static gauge. Because the coordinate transformation between static and light-cone gauge is field-dependent, the same correlation function in light-cone gauge involves operators evaluated at field-dependent locations $x_n^{lc}[\phi, x_n^s]$. These are complicated objects. For example calculating the correlation function as a path integral would involve operators that move around (in light-cone coordinates) as one integrates over the fundamental fields.

\section{Spectrum}
\label{sec5}

\subsection{Perturbation theory}

In this section we discuss the computation of the energy spectrum of our theory by applying ordinary perturbation theory\footnote{See also \cite{Lee:2021iut} for a similar discussion in the case of a $T\Tb$ deformed fermion.}.  This is useful in order to clarify how the $T\Tb$ trace relation resolves the quantization ambiguities.  Our Hamiltonian is \eqref{bm} 
where $\l=12\pi r_c/c_0$.  For simplicity we restrict here to the $b=\bb=0$ case, where the mode expansions \eqref{free field exp} and \eqref{free field exp 2} as well as the Hermiticity properties \eqref{eq:aDag} and \eqref{phi dagger} coincide. In terms of the fields $(f,\fb)$ \eqref{eq:f2phi}, which are both $2\pi$ periodic in $x$, we have the mode expansions
\eq{ab}{ f' & =  \sqrt{\frac{12}{ c_0}} \sum_{m\neq 0} \alpha_m e^{imx}~,\cr
 \fb' & = \sqrt{\frac{12}{ c_0}} \sum_{m\neq 0} \alphab_m e^{-imx}~, }
and impose the commutation relations \eqref{comutators}.
The Hamiltonian may be expanded to the first few orders as 
\eq{ad}{H = \frac{c_0}{ 24} \int_0^{2\pi} \frac{dx}{2\pi} \Big[ f'^2+\fb'^2 +\frac{1}{2}r_c f'^2 \fb'^2 +\frac{1}{8}r_c^2 f'^2 \fb'^2(f'^2+\fb'^2) + \ldots \Big] ~. }
 The general arguments of Zamolodchikov and Smirnov imply that we expect the energy spectrum to be
\eq{ae}{ E & = \frac{c_0 }{6r_c} \left[ 1-\sqrt{1-\frac{12 r_c}{c_0} (N+\Nb) +\frac{36 r_c^2}{ c_0^2}(N-\Nb)^2  }\right]~, \cr 
&= N+\Nb +\frac{12 r_c }{c_0}N\Nb +\frac{72 r_c^2 }{c_0^2} N\Nb(N+\Nb) + O\left( \frac{r_c^3} {c_0^3}\right) ~,  }
where $(N,\Nb)$ are eigenvalues of the  the level operators 
\eq{af}{ N= \sum_{m>1} \alpha_{-m}\alpha_m~,\quad  \Nb= \sum_{m>1} \alphab_{-m}\alphab_m~.}
To define $H$ as an operator and compute its spectrum we of course need to resolve  ordering ambiguities among the mode operators. We think of expanding  out $H$ in mode operators, order by order in $1/c_0$ (equivalently $r_c$).     Before reordering, the order $c_0^{-p}$ contribution to $H$ is a homogeneous degree $2p+2$ polynomial in the mode operators.  The difference between two choices of ordering  will involve a polynomial of degree $q<2p+2$ in the mode operators.    
Now imagine computing the matrix elements of $H$ in the unperturbed basis where states are labelled by occupation numbers in each mode, again expanding in  $1/c_0$.  Each squared matrix element at order $c_0^{-p}$ is a polynomial of degree  $2p+2$ in the occupation numbers.   The point we wish to emphasize here is that the order $2p+2$ piece of this polynomial is unambiguous in the sense that it is  independent of our choice of ordering and is also UV finite. For example, the difference between $\sum_{m>0} \alpha_{-m} \alpha_m $ and $\sum_{m>0} \alpha_{m} \alpha_{-m} $ is the UV divergent quantity $\sum_{m>1} m$, whose matrix elements are independent of the occupation numbers.   From \rf{ae} we see that the terms that appear in the energy spectrum involve only these unambiguous contributions.   We conclude that there is no ambiguity in reproducing these terms, and the only issue is how to fix terms that are subleading in $(N,\Nb)$ at a given order in the $1/c_0$ expansion.   

As we now discuss, these subleading ambiguities are fixed by the $T\Tb$ trace relation.  Classically, the relation
\eq{afa}{  T_{z\zb} = -\frac{6r_c }{c_0}( T_{zz} T_{\zb\zb} -   T_{z\zb}  T_{z\zb}  ) ~,}
is simply an identity.  However, we obtain a useful expression by taking the expectation value of both sides in an eigenstate of $H$ and $P$.   In particular, Zamolodchikov showed that the right hand side factorizes, in the sense that the following equation holds
\eq{ag}{ \langle  T_{z\zb} \rangle = -\frac{6r_c }{c_0}( \langle T_{zz} \rangle \langle T_{\zb\zb}\rangle -  \langle T_{z\zb} \rangle \langle T_{z\zb}\rangle  )~. }
Crucially, the expectation values  in this equation are independent of whatever improvement terms we wish to add to the stress tensor, as follows from the fact that improvement terms are derivatives of local operators, but local operators have constant expectation values in eigenstates of $H$ and $P$.  We are therefore free to impose \rf{ag} on the canonical stress tensor, which takes the form \eqref{ahz}
in terms of which the Hamiltonian \eqref{ad} is
\eq{aha}{ H =\int_0^{2\pi}  \frac{dx}{2\pi} T_{tt} =\int_0^{2\pi} \frac{dx}{2\pi} (2T_{z\zb}-T_{zz}-T_{\zb\zb})~.}
Our strategy is to carry out perturbation theory order by order in $1/c_0$, or equivalently in $r_c$, using \rf{ag} to fix any ordering ambiguities at each order, keeping in mind that both the states and the stress tensor receive corrections at each order.  We now illustrate this by working out the spectrum to order $1/c_0^2$, verifying the prediction of \rf{ae}. 

We start with the theory at $r_c=0$,   and define $(f'^2,\fb'^2)$ by normal ordering so that $E=N+\Nb$.  States are labelled by their eigenvalues under the individual level numbers \eqref{af}.  Now go to order $r_c$.   The trace relation tells us that $T_{z\zb} =-\frac{c_0}{96} r_c f'^2 \fb'^2$ with $(f'^2,\fb'^2)$ normal ordered.  

The Hamiltonian at order $r_c$ involves the expressions $f'^2 \fb'^2$ in $T_{zz}$ and $T_{\zb\zb}$.  It is natural to guess that these operators are also normal ordered, and we will indeed verify that is the correct prescription.  Indeed, with this assumption the correction to the energy is given by the standard formula (where subscripts indicate the order in $r_c$)
\eq{ai}{ E&  = E_0 + \langle \psi_0|H_1|\psi_0\rangle +\ldots  ~,\cr
& = N+\Nb + r_c \langle \psi_0 |\frac{c_0}{48}\int_0^{2\pi} \frac{dx }{2\pi}f'^2 \fb'^2|\psi_0\rangle +\ldots   ~,\cr
& = N+\Nb +\frac{12 r_c }{c_0}N\Nb+ \ldots ~,  }
in agreement with \rf{ae}.

At order $r_c^2$ the left hand side of \rf{ag} involves  $\langle \psi_0| f'^2 \fb'^2(f'^2 +\fb'^2)|\psi_0\rangle$, which is UV divergent.   We use \rf{ag} to define this matrix element in terms of matrix elements with fewer insertions of $(f,\fb)$.  At this order we also need to include corrections to the states according to the usual formula
\eq{aj}{ |E_n\rangle = |E_n^0\rangle + \sum_{k\neq n} \frac{ \langle E_k^0 |H_1| E_n^0\rangle}{ E_n^0 -E_k^0 } |E_k^0\rangle+ \ldots ~, }
where we note that since $H_1$ is diagonal within a degenerate subspace only non-degenerate unperturbed states contribute to the sum.

We can now work out the order $r_c^2$ correction to the energy from the standard formula of second order perturbation theory
\eq{ak}{ E_2 = \langle \psi_0 |H_2 |\psi_0\rangle + \sum_{\psi_0'\neq \psi_0} \frac{ |\langle \psi_0'| H_1|\psi_0\rangle|^2 }{ E_0-E_0'}~,}
with $H_2 = \frac{c_0}{24}\int_0^{2\pi}\frac{dx}{2\pi} \frac{1}{8} r_c^2 f'^2 \fb'^2(f'^2+\fb'^2) $.  We explained above how to relate the expectation value of $H_2$ to lower order matrix elements, but to see how the correct result comes out it is simpler to compute directly. More precisely, in order to confirm \rf{ae} we will just concentrate on the part cubic in occupation numbers, trusting that our general arguments enforce the vanishing of the terms subleading in occupation numbers. In terms of mode operators we have 
\eq{al}{H_2= \frac{9r_c^2}{ c_0^2} \sum_{m_i} (\alpha_{m_1}\alpha_{m_2}\alpha_{m_3}\alpha_{m_4}\alphab_{m_5}\alphab_{m_6} + \alpha_{m_1}\alpha_{m_2}\alphab_{m_3}\alphab_{m_4}\alphab_{m_5}\alphab_{m_6})~. }
The sum should be restricted such that the sum of $\alpha$ mode numbers equals the sum of $\alphab$  mode numbers.  The expectation value in a state with level numbers $\{N_m,\Nb_m\}$ works out to be 
\eq{am}{ \langle H_2 \rangle &=  \frac{108 r_c^2 }{c_0^2} \sum_{m_1,m_2} \left[ (N_{m_1})^2 \Nb_{m_2}+N_{m_1} (\Nb_{m_2})^2   \right]  \cr
&\quad +    \frac{432 r_c^2 }{c_0^2} \sum_{m_1,m_2,m_3 } \left[ N_{m_1} N_{m_2}  \Nb_{m_3}+N_{m_1} \Nb_{m_2}\Nb_{m_3}    \right]~.  }
On the other hand, the second order contribution in $H_1$ works out to be 
\eq{an}{ \sum_{\psi_0'\neq \psi_0} \frac{ |\langle \psi_0'| H_1|\psi_0\rangle|^2 }{ E_0-E_0'}&=-\frac{36 r_c^2 }{c_0^2} \sum_{m_1,m_2} \left[ (N_{m_1})^2 \Nb_{m_2}+N_{m_1} (\Nb_{m_2})^2   \right]\cr
& \quad -    \frac{288 r_c^2 }{c_0^2} \sum_{m_1,m_2,m_3 } \left[ N_{m_1} N_{m_2}  \Nb_{m_3}+N_{m_1} \Nb_{m_2}\Nb_{m_3}    \right]~.}
These nicely combine together in \rf{ak} to give 
\eq{ao}{ E_2 = \frac{72 r_c^2 }{c_0^2} N\Nb(N+\Nb)~,}
in agreement with \rf{ae}.  This procedure can in principle be carried out to higher orders.  As long as one enforces \rf{ag} at each order the result is guaranteed to reproduce \rf{ae}, since the trace relation \rf{ag} implies a differential equation for the energy levels that fixes their values, as shown by Zamolodchikov and Smirnov.   Our purpose in carrying out this exercise was to emphasize that the contributions that actually appear in \rf{ae} are insensitive to operator orderings.  Operator ordering ambiguities  only  affect terms of lower order in the level numbers, and these ambiguities are fixed by imposing the trace relation.  
\section{Correlation functions}
\label{sec6} 
\subsection{Gravity side}

For a Dirichlet boundary condition on the metric the standard choice of boundary term in \rf{Sdef} is \cite{Balasubramanian:1999re}
\eq{xa}{ S_{\rm bndy} = -\frac{1}{8\pi G} \int_{\partial M} \! d^2x \sqrt{h}(K-1) ~,}
where $h_{\mu\nu}$ is the metric on the boundary and $K$ denotes the trace of the extrinsic curvature.  We are setting the AdS$_3$ radius to unity,  $\ell=1$, for convenience.  The term without $K$ is not required by the variational principle but is added to yield a finite action in the limit that the boundary is taken to infinity.\footnote{Actually, finiteness also requires an additional term associated with the Weyl anomaly \cite{Henningson:1998gx}, but since this term is (locally) a total derivative it does not contribute to the stress tensor and will play no role in our discussion.}  The boundary stress tensor is defined via the on-shell variation of the action 
\eq{xb}{ \delta S = \frac{1}{4\pi} \int_{\partial M} \! d^2x \sqrt{h} T^{\mu\nu} \delta h_{\mu\nu}~,}
and works out to be 
\eq{xc}{T_{\mu\nu} = \frac{1}{4G}(K_{\mu\nu} - K h_{\mu\nu}+h_{\mu\nu})~.}

Now consider the problem of computing correlation functions of $T_{\mu\nu}$ on a flat boundary metric $h_{\mu\nu}$.  This boundary is taken to be a finite cutoff surface.   This can in principle be carried out order by order in $G \sim 1/c_0$.  The immediate question that arises is whether we will encounter UV divergences that require counterterms.  We first recall the standard argument that the  bulk Einstein-Hilbert action \rf{Sdef}  is not renormalized apart from a possible redefinition of $\ell$.  This follows from the fact that any candidate counterterm is proportional to $\int \! d^3x \sqrt{g}$ upon using the lowest order equation of motion; equivalently, a field redefinition may be used to eliminate any counterterm involving curvatures.\footnote{This statement is not without subtleties.  For example, if we employ dimensional regularization then away from $d=3$ there are independent curvature dependent counterterms.  These can potentially contribute as ``evanescent operators'', i.e., although the operators formally vanish as $\epsilon \rt 0$, if they are multiplied by $1/\epsilon$ coefficients they can still contribute in the limit.  We implicitly assume that there exists a renormalization scheme where such complications do not arise. }   

What about boundary terms?  The Gibbons-Hawking-York (GHY) term is tied by the variational principle to the Einstein-Hilbert term.  The boundary cosmological term can be fixed by the requirement that the stress tensor vanish on a flat planar boundary.  We should also consider boundary terms involving the intrinsic and extrinsic curvature.   A boundary term involving only intrinsic curvatures cannot contribute either to the action  or its first variation, and hence not to the stress tensor; recall here that we are interested in correlators on flat boundary.   Regarding extrinsic curvature terms, we demand that these do not spoil the variational principle.  The only type of boundary term which can contribute to the stress tensor without spoiling the variational principle takes the form 
\eq{xd}{ \Delta S_{\rm bndy} = -\frac{1}{4\pi}\int_{\partial M}\! d^2x \sqrt{h} R(h) Z(K_{\mu\nu})~,}
where $Z(K_{\mu\nu})$ can depend on the extrinsic curvature and boundary derivatives thereof.  This term contributes to the stress tensor on a flat boundary as an improvement term, 
\eq{xe}{ \Delta T_{\mu\nu} = (\pd_\mu \pd_\nu -  h_{\mu\nu}\pd^2 )Z(K)~.  }
 In the case of an asymptotic AdS$_3$ boundary \rf{xd} does not contribute since it dies off as the boundary is taken to infinity, but  it can contribute in the case of a finite cutoff.  
 
We conclude from this analysis that in the problem of computing stress tensor correlators on a flat boundary we can encounter two types of divergences: divergences in the action that vanish on-shell and hence can be removed by a field redefinition and divergences in the stress tensor corresponding to improvement terms.   This conclusion from the gravity side nicely matches our expectations on the 2d field theory side, as we now discuss.  
 
 \subsection{Field theory side}
The fact that on the gravity side all counterterms vanish for a flat boundary surface leads to the expectation that in field theory all counterterms in the action should vanish on-shell.  This implies that the S-matrix should be unambiguous, although this does not extend to  off-shell correlators of the elementary fields.   This expectation is supported by previous analyses \cite{Aharony:2010cx,Aharony:2010db,Dubovsky:2012sh,Dubovsky:2012wk,Aharony:2013ipa} of perturbation theory applied to the Nambu-Goto action\footnote{In  \cite{Dubovsky:2012sh,Dubovsky:2012wk} the Lagrangian form of the Nambu-Goto action was considered, rather than the Hamiltonian version that naturally arises in gravity, but we expect the conclusions to be the same.} and to $\TTb$ theories more generally, e.g \cite{Rosenhaus:2019utc}.     For example, the  one-loop  computation of the four-point function in  \cite{Ebert:2022cle} required the action counterterm $ \frac{r_c^2 }{4\pi \varepsilon} \int d^2x \pd_z(f'\bar{f}')\pd_{\zb}(f' \bar{f}')$.  This expression can be seen to vanish after integration by parts and using the lower order field equations $\pd_z \bar{f}' = \pd_{\zb} f'=0$.\footnote{This term is part of the evanescent 2d Ricci scalar that appears in dimensional regularization of the low energy effective action for long strings \cite{Dubovsky:2012sh}. We thank Raphael Flauger for bringing this to our attention.}   A compelling story regarding the finiteness of the $2\rightarrow 2$ S-matrix of the $d=3$ target space Nambu-Goto action in static gauge was laid out in   \cite{Dubovsky:2012sh,Dubovsky:2012wk,Dubovsky:2015zey,Dubovsky:2017cnj,Dubovsky:2018bmo}. 

We next turn to correlators of the stress tensor computed on the plane.  Previous work on this problem includes \cite{Kraus:2018xrn,Aharony:2018vux,Cardy:2019qao,Hirano:2020ppu,Rosenhaus:2019utc,Dey:2020gwm,He:2020udl,Ebert:2022cle}.  We first recall the renormalization properties of the stress tensor in a general QFT \cite{Brown:1979pq}.  Let $S(\phi)$ be the renormalized action; that is $S(\phi)$ includes counterterms that render finite all correlators of the elementary $\phi$ fields.   Let $T_{\mu\nu}$ be the canonical stress tensor \eqref{can T} obtained from $S(\phi)$ by application of Noether's theorem.  The stress tensor Ward identity relates correlators of $\pd^\mu T_{\mu\nu}$ with elementary fields to correlators of elementary fields alone.  Since the latter are assumed to be finite, this implies that $\pd^\mu T_{\mu\nu}$ is a finite operator.   The only divergences that can show up in $T_{\mu\nu}$ therefore take the form of identically conserved tensors\footnote{By ``identically conserved'' we mean that the conservation should hold without using the equations of motion, since an equation of motion piece would spoil the contact terms in the Ward identity.}, which for a scalar field theory take the form of $\Delta T_{\mu\nu} = (\pd_\mu \pd_\nu - \eta_{\mu\nu} \pd^2)Y(\phi)$, which we refer to as an improvement term.  Besides the improvement terms needed for finiteness there is also the freedom to add finite improvement terms, and these must enter into the discussion of what one means by stress tensor correlators.  

In a CFT tracelessness of the stress tensor restricts the class of improvement terms that can appear, and indeed Virasoro symmetry fixes the stress tensor correlators uniquely in terms of a single number, the central charge $c$.  In a $\TTb$ deformed CFT the stress tensor is not traceless; at the classical level it obeys the trace equation $ \Tr T =\frac{\lambda}{2\pi} \det T$, and we have seen how this equation may be used to fix the form of the stress tensor order by order in $\lambda$ starting from an initial seed.  It is natural to ask to what extent the quantum version of the trace equation fixes the form of the quantum stress tensor and its correlators.

To address this we need to discuss the definition of the composite operator $\det T$ in the quantum theory.  Zamolodchikov famously showed that this operator is well defined (UV finite) up to total derivatives of local operators \cite{Zamolodchikov:2004ce}.  This has the important consequence that the expectation value of $\det T$ is any energy-momentum eigenstate is finite and unambiguous, since the total derivative pieces vanish in such states by translational invariance.  However, the  total derivatives terms do contribute to off-diagonal matrix elements, so in general we have to accept that $\det T$ requires such terms in order to make sense as an operator. 

Now suppose that we have defined a renormalized stress tensor that obeys the relation
\eq{ga}{ \Tr T = \frac{\lambda}{2\pi} \det T + \pd_\mu W^\mu~, }
where $\pd_\mu W^\mu$ denote operators added to   $\det T$ to achieve UV finiteness.   Then, consider adding an infinitesimal improvement term to the stress tensor, $T'_{\mu\nu} = T_{\mu\nu}+ (\pd_\mu \pd_\nu -\eta_{\mu\nu}\pd^2)Y $.  A simple computation shows that the new stress tensor obeys the equation $ \Tr T' = \frac{\lambda}{2\pi} \det T' + \pd_\mu W'^\mu$ with $W'^\mu = W^\mu -2 \pd^\mu Y -\frac{\lambda }{2\pi} T^{\mu\nu} \pd_\nu Y$, where we used conservation of the stress tensor.  The new stress tensor therefore satisfies a trace equation that is equally valid as the original one.  We conclude that the quantum trace equation does not fix the form of the quantum stress tensor, even though it does so at the classical level, the reason being the appearance of the $\pd_\mu W^\mu$ term that is needed for finiteness.  It follows that stress tensor correlators are similarly ambiguous.  It is possible that there is some additional principle that may be imposed on the stress tensor to render it unambiguous, but this would presumably require a better understanding of how to define a $\TTb$ deformed theory at the non-perturbative level.  

We end this section by noting how these conclusions are borne out in perturbation theory.  In \cite{Ebert:2022cle} the two-point function  $\langle T_{z\bar{z}} T_{z\bar{z} }\rangle $ was computed to two-loop order; all other stress tensor two-point functions may be obtained from this by Ward identities.    In dimensional regularization the stress tensor was found to require the counterterm $\Delta T_{z\bar{z}} = -\frac{1}{2\varepsilon}r_c^2 f'''  \bar{f}''' =-\frac{1}{2\varepsilon}r_c^2  \pd_z \pd_{\bar{z}} (f'' \bar{f}'') $, where we used the equations of motion in the last line in order to define the counterterm as an improvement term.  The addition of this improvement term then induces the appearance of a $W^\mu$ term in the trace equation.  This pattern is expected to persist at all orders in perturbation theory. 

\section*{Acknowledgments}

The research of PK and RM is supported in part by the National Science Foundation under research grant PHY-19-14412. The work of KR is supported by the Mani L. Bhaumik Institute for Theoretical Physics
\appendix
\section{Deformed action from the trace equation}%
\label{appA}
In this appendix we show how one can determine the deformed action starting from the trace equation \eqref{eq:Tr} in the first order formalism using $\phi$ and its canonical momentum $\Pi$. Although this approach might be most appropriate for this work since we are mostly working in the first order formalism which makes quantization easier, it is not very convenient for obtaining closed form results mainly because it obscures Lorentz invariance. 

We illustrate how one can construct the deformed action as a power in series in $\l$ and we determine the first order correction. Let's recall the definition of the canonical stress tensor 

\begin{equation}
    T_{\m\n} = 2\pi\left(\d_{\m\n}L-\frac{\pd L}{\pd (\pd^\m \phi)} \pd_\n \phi\right)~.
\end{equation}
Using the definition of the canonical momentum 
\begin{equation}
    \Pi = \frac{\pd L}{\pd \dot{\phi}}~,
\end{equation}
we can express the components of the stress tensor as 
\begin{equation}
-\frac{1}{2\pi}T_{00}= H, \quad -\frac{1}{2\pi}T_{01}= \Pi\phi', \quad  -\frac{1}{2\pi}T_{10}= \frac{\pd H}{\pd \Pi} \frac{\pd H}{\pd \phi'}, \quad -\frac{1}{2\pi}T_{11}=H-\frac{\partial H}{\partial \Pi} \Pi -\frac{\partial H}{\partial \f'} \f'~. \label{T comps in pi}
\end{equation}

To first order the most general gauge invariant (hence no $\f$ polynomials) expression for the Hamiltonian density is
\begin{equation}
    H = \frac12 (\f^{'2} + \Pi^2) + \l( c_1 \f^{'4} + c_2 \f^{'3} \Pi+ c_3 \f^{'2} \Pi^2+ c_4 \f^{'} \Pi^3+c_5 \Pi^4)+ \mathcal{O}(\l^2)~.
\end{equation}
In this approach Lorentz invariance is encoded in the symmetry of the stress tensor 
\begin{equation}
 T_{01}=T_{10} \Rightarrow \Pi \f' = \frac{\partial H}{\partial \Pi}\frac{\partial H}{\partial \f'~.}
\end{equation}
This fixes $c_2=c_4=0,c_1=c_5$ and $c_3=-2c_5$. H
ence only one coefficient remains by just requiring Lorentz invariance. Using \eqref{T comps in pi} we then find

\begin{align}
    -\frac{1}{2\pi}T_{00} &= \frac12 (\f^{'2} + \Pi^2) + c_5 \l(\f^{'2} - \Pi^2)^2+ \mathcal{O}(\l^2)~, \\
    -\frac{1}{2\pi}T_{01} &= T_{10} =  \Pi\phi' ~,\\
    -\frac{1}{2\pi}T_{11} &=  -\frac12 (\f^{'2} + \Pi^2) - 3c_5 \l (\f^{'2} - \Pi^2)^2+ \mathcal{O}(\l^2)~.
\end{align}
Plugging the above into the trace equation \eqref{eq:Tr} we can fix the reaming coefficient  $c_5=\frac{1}{8}$.

\section{Conversion to Lorentzian signature}
\label{appLor}
We relate the Euclidean and Lorentzian times as  
\eq{za}{ t_E = -i t_L~. } 
The (anti)-holomorphic coordinates then become light-cone coordinates
\eq{zb}{ z = x+it_E = x+t_L = x^+ ~, \cr
\zb = x-it_E = x-t_L = x^- ~,}
so 
\eq{zca}{ \pd_+ = \frac{1}{2}( \pd_x + \pd_{t_L})~,\quad  \pd_- = \frac{1}{2}( \pd_x - \pd_{t_L})~. }
Since $dt_E = -idt_L$ the relation between the actions is $S_E = i S_L$, i.e. so that 
\eq{zd}{ S_E = \int dt_E dx( (\pd_{t_E} \phi)^2 + \phi'^2) = -i \int dt_L dx( -(\pd_{t_L} \phi)^2 + \phi'^2) =i S_L~.   }
The Euclidean action \eqref{Free ffb} then converts to the Lorentzian action 
\eq{zf}{ S_L =  \frac{c_0}{24\pi} \int\! dx dt_L ( f' \pd_- f + \fb'\pd_+ \fb)~.} 
We keep the relation between $f, \fb$ and $\phi, \Pi$ as in \eqref{eq:f2phi} and find
\eq{zh}{ S_L =  \int\! dx dt_L \big( \Pi \dot{\phi} -\frac{1}{2}( \phi'^2 + \Pi^2) \big)~.}

\bibliographystyle{jhep}
\bibliography{refs}
\end{document}